\documentclass[aps,english,twocolumn,prb,superscriptaddress]{revtex4}
\usepackage{graphicx}
\usepackage{amssymb}
\usepackage{babel}

\begin{document}
\title{Accurate projected augmented wave (PAW) datasets for rare-earth elements (RE=La-Lu)}
\author{M. Topsakal}
\affiliation{Department of Chemical Engineering and Materials Science, University of Minnesota, Minneapolis, MN, USA} 
\author{R.M. Wentzcovitch}
\affiliation{Department of Chemical Engineering and Materials Science, University of Minnesota, Minneapolis, MN, USA}

\date{\today}

\begin{abstract}
We provide accurate projected augmented wave (PAW) datasets for rare-earth (RE) elements with some suggested Hubbard U values allowing efficient plane-wave calculations. Solid state tests of generated datasets were performed on rare-earth nitrides. Through density of state (DOS) and equation of state (EoS) comparisons, generated datasets were shown to yield excellent results comparable to highly accurate all-electron full-potential linearized augmented plane-wave plus local orbital (FLAPW+LO) calculations. Hubbard U values for trivalent RE ions are determined according to hybrid functional calculations. We believe that these new and open-source PAW datasets will allow further studies on rare-earth materials.
\end{abstract}

\maketitle

\section{Introduction} \label{sec:intro}

The lanthanide series of the periodic table comprises fifteen members ranging from 
Lanthanum (La) to Lutetium (Lu). Although they are more abundant than silver, and 
some of them are more abundant than lead, they are known as rare-earth (RE) elements. The ``rare`` in 
their name refers to the difficulty of obtaining the pure elements, not to their abundances in nature. 
They are never found as free metals in the Earth's crust and do not exist as pure minerals. All having two electrons in the outermost 6\textit{s} orbital, they can form trivalent cations. Their chemistry is largely determined by the ionic radius, which decreases steadily along the series corresponding to the fillings of the 4\textit{f}-orbitals.

One of the principal applications of the rare-earth elements in industry, involving
millions of tons of raw material each year, is in the production of catalysts
for the cracking of crude petroleum.\cite{book1} They are 
also commonly used in the glass and display 
industry.\cite{book2} Additionally, rare-earth oxides are regarded as potential candidates 
for high-k gate dielectrics because of their ability to 
form direct contact with silicon substrate.\cite{sinisa} Superconductors based on rare-earth oxypnictides\cite{sc1,sc2,sc3} were also discovered with critical temperatures as high as 55 K.\cite{sc4} Very recently
it was shown that the rare-earth oxide ceramics are intrinsically 
hydrophobic and durable materials preventing water 
from spreading over a surface.\cite{mit-hydro} S. Bertaina \textit{et al.}\cite{qubit} 
revealed a new family of spin qubits based on RE ions having desired characteristics suitable for scalable quantum 
computation at $^4$He temperatures. Adsorption of RE adatoms on graphene were investigated by several studies. 
In contrast to most of simple and transition metals, it was shown that RE adatoms induce significant electric dipole and 
magnetic moments on graphene.\cite{ads1,ads2,ads3}

Although RE elements constitute a significant portion of atoms in the periodic table and exhibit unique (and mostly unexplored)  properties, it is interesting to note that the ab-initio electronic structure calculations with RE elements are very scarce. This is due to the fact that 
RE elements containing 4\textit{f}-electrons are particularly very challenging for density functional theory (DFT) methods. It is well known that the narrow \textit{f}-bands in RE compounds are not adequately described by standard local density approximation (LDA)\cite{lda} and generalized-gradient approximation (GGA)\cite{gga} due to strong electronic correlation effects. To cope with this insufficiencies, the DFT+U method\cite{dftpu} is often employed, with  U fit to some spectral data. Another approach is to use a hybrid functional introducing a portion of exact Hartree-Fock type exchange into the exchange-correlation functional.\cite{hse,kresse_ce,hse_ce} Both approaches improve the treatment of strongly correlated electrons. However DFT+U suffers from ambiguity of Hubbard U value while hybrid functionals suffer from extremely demanding computational costs. 

We realized that reliable DFT potentials for majority of RE elements are not provided by any open-source\cite{vasp} DFT simulation environments such as ABINIT,\cite{abinit} SIESTA, \cite{siesta} and QUANTUM ESPRESSO.\cite{qe,gbrv}  The aim of this study is to provide projected augmented wave (PAW) potentials for RE elements with some suggested Hubbard U values allowing efficient plane-wave calculations. These potentials were developed to yield results comparable to highly accurate all-electron FLAPW+LO results using WIEN2K.\cite{wien2k} Hubbard U values are consistently determined according to all-electron hybrid functional calculations. The optimization of the potentials were performed on rare-earth nitrides in which RE element adopts trivalent ionic state.

The paper is organized as follows: In section \ref{sec:method} we present the details about the generation of RE PAW datasets. Next we investigate the electronic properties of RE-nitrides (REN's) which will be used as a reference for validation of our generated PAW datasets. In section \ref{sec:comp} we compare the performance of generated PAW datasets with all-electron FLAPW+LO results. In section \ref{sec:match} we present our procedure of obtaining Hubbard U values for feasible DFT+U calculations that match results from demanding hybrid functional (YS-PBE0) calculations. Our conclusions are summarized in section \ref{sec:conc}.

\section{Methods and calculation details} \label{sec:method}

We used ATOMPAW program\cite{atompaw1,atompaw2,atompaw3} to generate projector and basis functions which are needed for performing 
first principles calculations based on the projector augmented wave (PAW) method.\cite{paw} For any element in the periodic table, the program inputs the atomic number, exchange-correlation functional, electronic configuration, choice of basis functions, and a cutoff radii. After initialization, the program can be instructed to output potential files which can be used by open-source density functional simulation environments such as QUANTUM ESPRESSO\cite{qe,qe2} and ABINIT.\cite{abinit,abinit2} It also outputs several files enabling wave function and logarithmic derivative plots.

Although Scandium (Sc) and Yttrium (Y) belong to rare-earth element family, we focus our attention in the lanthanides (La-Lu) 
since PAW datasets already exist for Sc and Y in the ATOMPAW repository. For each lanthanide we used neutral electronic configuration. 6\textit{s}, 5\textit{s}, 5\textit{p}, 5\textit{d} and 4\textit{f} orbitals are treated as valance states. Two projector and basis functions were used for each angular momentum channel (s,p,d,f). Two Rydberg reference energy was used to build each partial-waves. Since lanthanides are high-Z elements, scalar-relativistic wave equation was used for all-electron computations. VANDERBILT scheme\cite{vanderbilt_scheme} was used to pseudize partial-waves. The analytical form of shape functions used in compensation density was chosen according to the BESSELSHAPE option.\cite{besselshape} The generated dataset was carefully checked with logarithmic derivatives against ghost states. [Xe] 4\textit{f}$^{(n)}$5d$^{1}$6s$^{2}$ atomic configuration was assumed for each RE in which n=0 for La, n=1 for Ce, n=2 for Pr, n=3 for Nd, etc. 

Cutoff radius\cite{atompaw1} (r$_{cut}$) of augmentation regions in the PAW formalism was determined according to solid state calculations performed on RE-nitrides (REN's). We first obtained equilibrium lattice constants (a$_0$) for each REN using the all-electron WIEN2K\cite{wien2k,wien2k2} program which employs highly accurate full-potential linearized augmented plane-wave + local orbitals (FLAPW+LO) method. Then we tuned cutoff radius of each PAW dataset to yield equilibrium lattice constant error not larger than $\pm$0.03 \% relative to WIEN2K. Table \ref{table:radius} lists r$_{cut}$ values for each RE.

\begin{table}
\caption{Muffin-tin radii values (\textit{R}$_{MT}$) used in WIEN2K calculations and cutoff radius (r$_{cut}$) of augmentation regions in PAW method used in ATOMPAW.}
\label{table:radius} \centering{}
\begin{tabular}{|c|c|c|c|}
\hline
Element & \textit{R}$_{MT}$ (a.u)  & r$_{cut}$ (a.u) 
\tabularnewline
\hline
\hline
La &   2.470  & 2.450  \tabularnewline               
\hline                                         
Ce &   2.350  & 2.335  \tabularnewline               
\hline                                        
Pr &   2.410  & 2.490  \tabularnewline               
\hline                                        
Nd &   2.430  & 2.450  \tabularnewline               
\hline                                       
Pm &   2.400  & 2.450  \tabularnewline               
\hline                                        
Sm &   2.390  & 2.516  \tabularnewline               
\hline                                       
Eu &   2.400  & 2.559  \tabularnewline               
\hline                                         
Gd &   2.370  & 2.417  \tabularnewline               
\hline                                        
Tb &   2.330  & 2.370  \tabularnewline               
\hline                                        
Dy &   2.260  & 2.392  \tabularnewline               
\hline                                        
Ho &   2.320  & 2.370  \tabularnewline               
\hline                                        
Er &   2.310  & 2.375  \tabularnewline               
\hline                                       
Tm &   2.290  & 2.393  \tabularnewline               
\hline                                       
Yb &   2.280  & 2.400  \tabularnewline               
\hline                                          
Lu &   2.270  & 2.250  \tabularnewline               
\hline                               
\hline
\end{tabular}
\end{table}

In all-electron (WIEN2K) calculations,  the wave functions were expanded in spherical harmonics inside non-overlapping atomic spheres of radius \textit{R}$_{MT}$ (muffin-tin radii), and in plane waves for the remaining space of the unit cell. WIEN2K \textit{R}$_{MT}$ defaults of each rare-earth and nitrogen (N) atoms are reduced by \%10. Corresponding \textit{R}$_{MT}$ values of each RE are presented in Table \ref{table:radius}.  The plane wave expansion in the interstitial region was made up to a cut-off wave vector chosen to be \textit{K}$_{max}$ = 9.0/\textit{R}$_{MT}$ which is \%28 larger than the WIEN2K default. Setting \textit{K}$_{max}$ to 10.0/\textit{R}$_{MT}$ or 11.0/\textit{R}$_{MT}$ did not change equation of state results noticeably while significantly increasing computational time. The Brillouin-zone is sampled at 12x12x12 k-points using the tetrahedron method. To improve the visibility, density of states (DOS) plots were smoothed using a Gaussian with 0.003 Ry width. Hybrid functional calculations are performed using the YS-PBE0 functional (where YS stands for Yukawa screened) as implemented in WIEN2K. This functional was shown to yield similar results\cite{wien2k-hse} to the original HSE functional which is a screened hybrid functional.\cite{hse} One quarter of the PBE\cite{pbe} short-range exchange is replaced by the exact exchange, while the full PBE correlation energy is included. The hybrid calculations were carried out using a 9x9x9 k-points mesh. For DFT+U calculations, the standard Duradev implementation\cite{duradev} is used where onsite Coulomb interaction for localized orbitals is parametrized by U$_{effective}$ = \textit{U - J} ( which we denote henceforth as \textit{U$_{d}$} and \textit{U$_{f}$} for \textit{d} and \textit{f} orbitals ). We select \textit{J} = 0.
 
For verification of PAW datasets, we performed plane-wave calculations using QUANTUM ESPRESSO. K-points sampling, smearing parameters and DFT+U method was chosen similar to WIEN2K calculations while 50 (200) Ry wave function (charge density) cutoff was used. N.pbe-kjpaw.UPF potential (provided by QUANTUM ESPRESSO web page\cite{qe-web}) was used for Nitrogen atom. In all WIEN2K and QUANTUM ESPRESSO calculations, spin polarized PBE exchange correlation functional was used.

For the rest of the paper, we shall simply refer to rare-earth as ''RE'', rare-earth nitrides as ``REN's``, all-electron FLAPW+LO as ''AE``, plane augmented wave dataset as PAW, calculations with GGA-PBE functional as ``PBE``, hybrid functional calculations as ''YS-PBE0`` when applicable.

\begin{figure*}
\includegraphics[width=18cm]{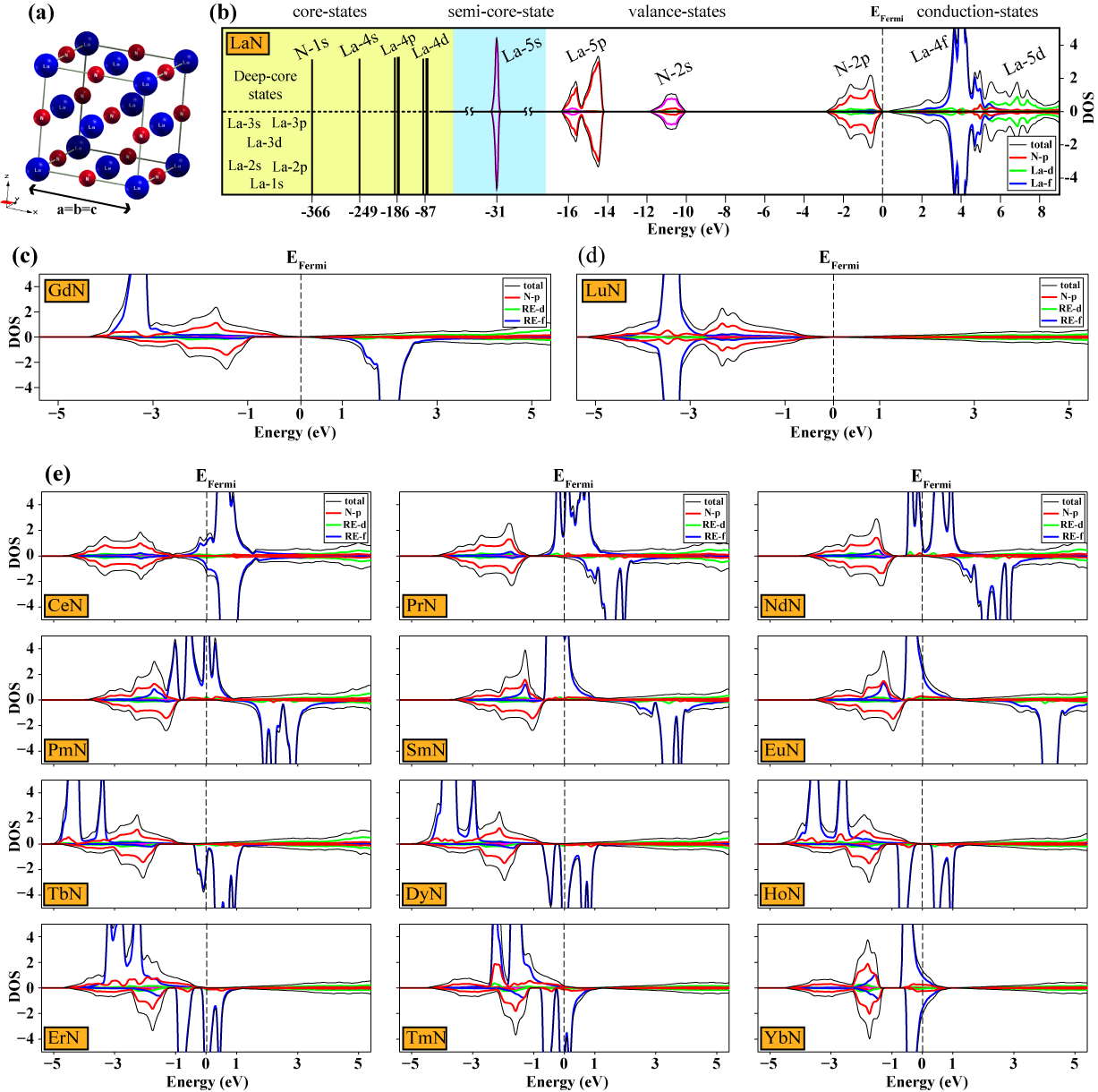}
\caption{(Color online) Cubic unit cell of LaN. (b) Total and projected density of states (DOS) of LaN calculated using all-electron FLAPW+LO method. PBE exchange-correlation functional was used. Core-states are shaded. Positive (negative) y-axis corresponds to DOS of spin-up (spin-down) states. Zero of the energy was set to Fermi energy (E$_{Fermi}$). (c) DOS of GdN having seven f-orbitals filled and seven f-orbitals empty with total magnetic moment $\mu$=7 $\mu$B per primitive cell. (d) DOS of LuN having completely filled f-orbitals. (e) DOS for other REN's calculated using all-electron FLAPW+LO method.}
\label{fig:000}
\end{figure*}

\section{AE calculations on rare-earth nitrides} \label{sec:wien}
The fidelity of the logarithmic derivatives of the pseudo wave functions in comparison with
their all-electron counterparts is a requirement for developing a reliable PAW dataset. However it is by no means sufficient. One needs to test the performance of generated dataset in realistic environments through solid-state calculations. In this section, we try to build our understanding of the electronic properties of rare-earth elements through reliable all-electron WIEN2K FLAPW+LO calculations to which we will frequently refer in the following sections. In a similar manner, Lejaeghere \textit{et al.}\cite{lajaeghere} have used WIEN2K as a reference to describe quantitatively the discrepancies in the equation of state of a wide set of elements in the periodic table (71 elements, excluding lanthanides). We also believe that our all-electron calculations in this section might provide useful information for other studies involving RE-elements.

We select rare-earth nitrides (REN's) as testing material due to their simple rocksalt structure and trivalent state of RE. As discussed in the introduction section, RE elements are never found as free metals in the Earth's crust and do not exist as pure minerals. They are very reactive with oxygen in the ambient atmosphere. Generally the trivalent state (RE$^{+3}$) is the most stable form and therefore, sesquioxides\cite{re2o3} (RE$_2O_3$), nitrides\cite{ReN-exp} (REN), cobaltates\cite{ReCoO3} (RECoO$_3$) exist for  rare-earth elements. We believe that our generated RE datasets should also work equally well for other RE ions since they are generated from neutral electronic configuration.

The atomic structure of LaN is illustrated In Fig. ~\ref{fig:000} (a). The crystal system is cubic and belongs to the (Fm$\bar{3}$m) symmetry group. Unless stated otherwise, we use experimental lattice constants\cite{ReN-exp,Pm} for any DOS and band structure calculation in this study. Total and local density of states for LaN is presented in Fig. ~\ref{fig:000} (b) calculated using standard PBE functional. Valance states near the Fermi level have mainly Nitrogen-2\textit{p} character while Nitrogen-2\textit{s} states lie approximately 11 eV below the Fermi level. The La-5\textit{d} and La-6\textit{s} orbitals don't have significant contribution in the valance states since they are ionized. The La-5\textit{s} semi-core state is approximately 31 eV below the Fermi level and lower lying states of La (4\textit{d}, 4\textit{p}, 4\textit{s}...)  and N-1\textit{s} orbitals are considered as core-states as depicted in Fig. ~\ref{fig:000} (b). Empty La-4\textit{f} states lie approximately 4 eV above the Fermi level. LaN is a nonmagnetic semi-metal and its PBE band structure is shown in ~\ref{fig:hse:band} (b). Lowest energy conduction states have La-5\textit{d} orbital character.

Starting from Cerium (Ce), 4\textit{f} orbitals start to get occupied in the REN's. Gadolinium nitride (GdN) is the case in which seven 4\textit{f} orbitals are half occupied so it has a huge magnetic moment as $\mu$=7 $\mu_B$ per primitive cell. Total and local density of states for GdN are presented in Fig. ~\ref{fig:000} (c). Similar to LaN, the top of valance band has primarily N-\textit{p} states. Half occupied 4\textit{f} states are approximately 3.2 eV below the Fermi level and they interact with N-\textit{p} states. For the case of Lutetium nitride (LuN), each of f-orbitals are fully occupied so we have a nonmagnetic case. Total and local density of states for LuN is presented in Fig. ~\ref{fig:000} (d).

Fig. ~\ref{fig:000} (e) shows total and local density of states for the rest of REN's. One can easily track the increase of occupied spin-up electrons from Ce to Eu and spin-down electrons from Tb to Yb. In CeN, the single f-electron is distributed in spin-up and spin-down orbitals and it produces $\mu$=0.27 $\mu_B$ per primitive cell. Similarly YbN has only one empty 4\textit{f} orbital which is distributed in spin-up and spin-down orbitals and it produces $\mu$=0.22 $\mu_B$ per primitive cell. Others have almost integer magnetic moments, which are $\mu$=1.97, 3.00, 4.00, 5.00, 6.00, 6.03, 5.01, 4.00, 3.00, 1.93 $\mu_B$ for PrN, NdN, PmN, SmN, EuN, TbN, DyN, HoN, ErN, TmN respectively. Different from LaN, GdN, and LuN, the energies of f-states are very close to the Fermi level for REN's as shown in Fig. ~\ref{fig:000} (e).

\section{Tests of generated PAW datasets} \label{sec:comp}

\begin{figure*}
\includegraphics[width=18cm]{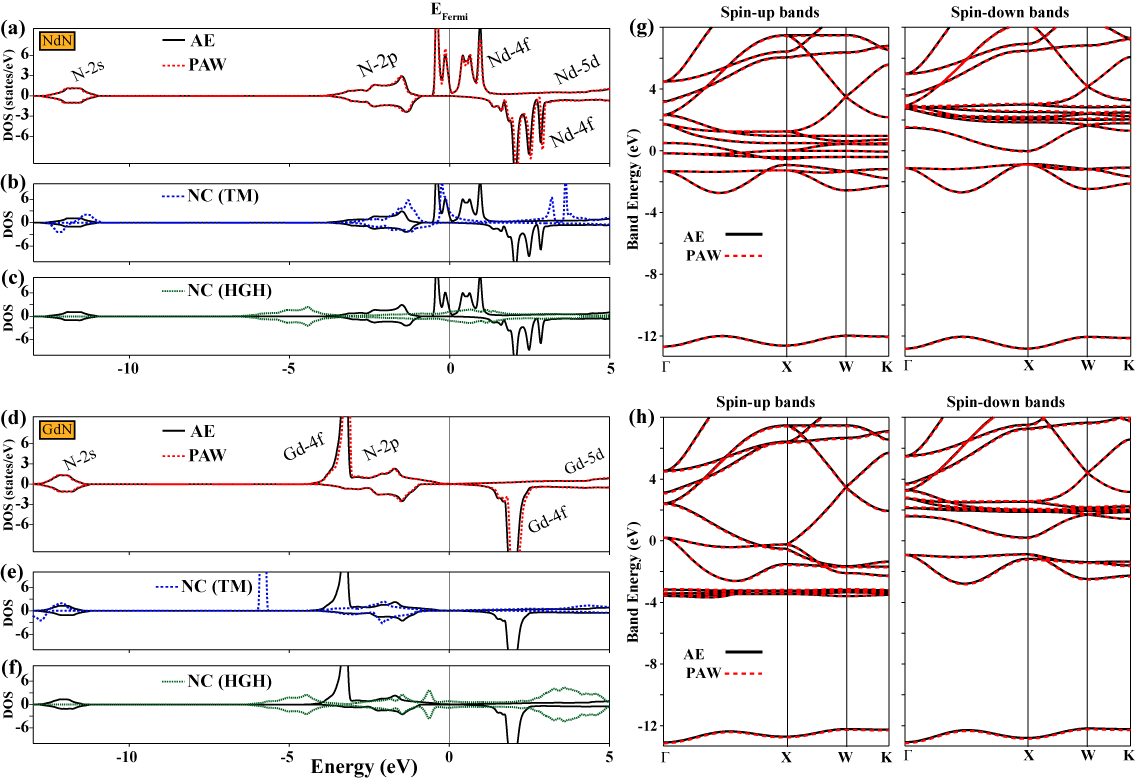}
\caption{(Color online) (a) Density of states (DOS) comparison of AE and PAW calculations for NdN. Positive (negative) y-axis corresponds to DOS of spin-up (spin-down) states. Zero of the energy was set to Fermi energy (E$_{Fermi}$). PBE exchange-correlation functional was used. (b) DOS comparison of AE and Troullier-Martins (TM) norm-conserving (NC) pseudopotential calculations for NdN. (c) DOS comparison of AE and Hartwigsen-Goedecker-Hutter (HGH) norm-conserving (NC) pseudopotential calculations for NdN. (d) Same as (a) for GdN. (e) Same as (b) for GdN. (f) Same as (c) for GdN. (g) Band structure comparison of AE and PAW calculations for NdN. (h) Same as (g) for GdN. }
\label{fig:comp}
\end{figure*}

\begin{figure}
\includegraphics[width=8cm]{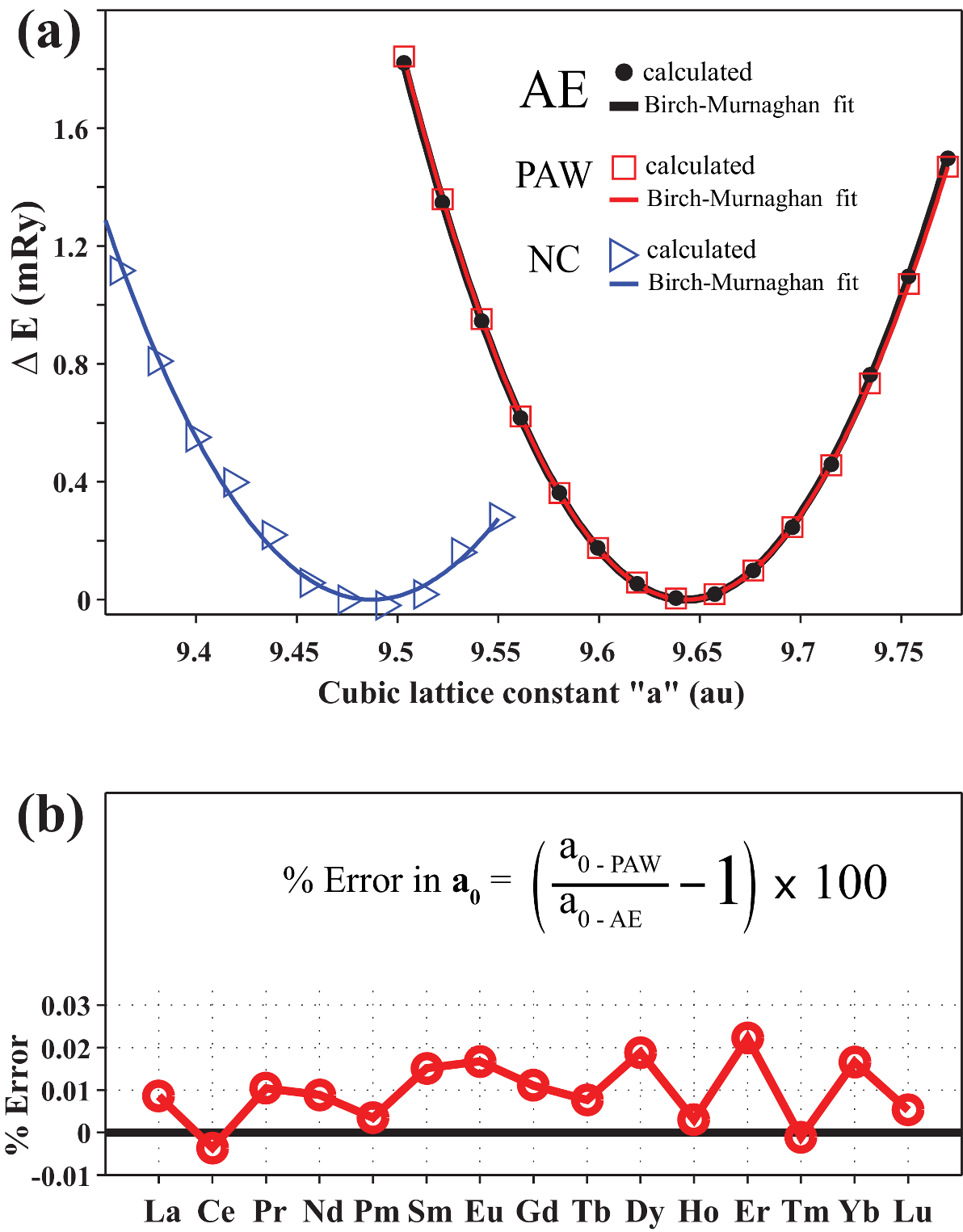}
\caption{ (Color online) (a) Equation of state (EoS) comparison of AE, PAW and NC calculations for NdN. Minimum of the energy was set to zero for each calculation. (b) Percent difference in AE versus PAW calculations for REN equilibrium lattice constants (a$_0$). }
\label{fig:eos}
\end{figure}

In this section we test our generated PAW datasets with respect to their AE counterparts as presented in the previous section. For each RE, generated PAW datasets were converted into UPF format\cite{upf} and solid state calculations were performed using the QUANTUM ESPRESSO (QE). However it is also possible to perform these calculations using ABINIT since we also provide each PAW dataset in .abinit format readable by the ABINIT. Lattice constants, atomic positions, k-point sets and exchange-correlation functional parameters are kept identical for AE and PAW calculations.

 The basic requirement is the matching of PAW total DOS with AE total DOS. Since we are using completely different software packages to get the DOS spectrum, very small differences are tolerable. In Fig. ~\ref{fig:comp} (a) we compare total DOS of NdN and the agreement is fairly good. PAW (dashed) lines  mostly overlap with AE (continuous) lines in the DOS. We believe that it is worthwhile to make similar comparison with previously generated RE pseudopotentials which are available on QUANTUM ESPRESSO repository\cite{qe-web} in order to reveal the importance of RE potentials on solid state DFT calculations. Nd.pbe-mt\_fhi.UPF is a norm-conserving (NC) type pseudopotential generated with Troullier-Martins (TM) method.\cite{TM} In Fig. ~\ref{fig:comp} (b) total density of states obtained with this pseudopotential is compared with AE DOS. As apparently seen, there is a significant disagreement between NC and AE density of states - especially for f-states. In Fig. ~\ref{fig:comp} (c), we compare another NC pseudopotential (Nd.pz-sp-hgh.UPF) which was generated using Hartwigsen-Goedecker-Hutter method (HGH).\cite{HGH} Again the disagreement is quite pronounced. HGH potential even fails to produce a spin-polarized solution in contrast to PAW and TM results. This might be related with placement of f-electrons in the core region or failure of generated NC pseudopotential in trivalent ionic state. For GdN, generated PAW dataset agree excellently with AE as shown in Fig. ~\ref{fig:comp} (d). On the other  TM and HGH pseudopotentials ( Gd.pbe-mt\_fhi.UPF, Gd.pz-sp-hgh.UPF ) significantly fail as presented in Fig. ~\ref{fig:comp} (e) and (f). All these results indicate that the choice of potentials are extremely important in DFT calculations. 

In Fig. ~\ref{fig:comp} (g) and (h) we compare the PBE band structure of NdN and GdN calculated with AE and PAW. Similar to DOS comparisons, band structures are almost identical between AE and PAW calculations for spin-up and spin-down bands. For the rest of REN's, results of DOS and band structure tests are presented in Supplementary materials (see section \ref{sec:appendix}) each showing excellent agreement with AE.

Upon achieving our first requirement that PAW total DOS and band structure should practically match with AE where each calculation was performed at experimental lattice constant, our next requirement is that similar agreement should also happen for different volumes where distances between atoms change. Equation of state (EoS) calculation is a way of performing this test. The agreement of EoS for a generated potential with its all-electron counterpart is another measure of its reliability. For each REN's we have obtained their EoS calculated with AE and PAW. 

Our strategy for EoS tests is as follows: First we obtain the AE energy versus volume curve for all REN’s as exemplified in Fig. ~\ref{fig:eos} (a). Results are fitted to a third order Birch-Murnaghan equation of state\cite{B-M} 

\begin{widetext}
\begin{equation}
E(V)=E_{0}+\frac{9}{16}B_{0}V_{0}\left \{  \left [ \left (\frac{V_{0}}{V} \right )^{2/3}-1 \right ]^{3} B_{0}{}'+ \left[ \left(\frac{V_{0}}{V} \right )^{2/3}-1\right ]^{2}\left [ 6-4 \left(\frac{V_{0}}{V} \right)^{2/3} \right ]  \right \}  ,
\end{equation}
\end{widetext}

where E$_0$ is the equilibrium total energy, V$_0$ is the equilibrium volume, and B$_0$ and B$_0{}'$ are the bulk modulus and its pressure derivative, respectively. The corresponding cubic lattice constant (a$_0$) is obtained from V$_0$. We repeat this procedure with QE PAW calculations and energy minima of each curve are set to zero. If AE and PAW V$_0$ values do not agree well, we change the cutoff (r$_{cut}$) of RE and generate a new PAW dataset. This procedure is repeated until the difference ($\%$ error) between AE and PAW lattice constants (a$_0$) are less than $\pm$0.03\%. In Fig. ~\ref{fig:eos} (a) AE and PAW EoS's are compared for NdN. Apparently, AE and PAW results are almost identical, similar to DOS and band structure tests presented in Fig. ~\ref{fig:comp} (a). For sake of comparison, EoS of NdN calculated with Troullier-Martins (TM) norm-conserving (NC) type pseudopotential\cite{qe-web} is also presented in Fig. ~\ref{fig:eos} (a). Since the electronic structure of NdN calculated using NC pseudopotentials is significantly different from PAW and AE calculations as shown in Fig. ~\ref{fig:comp} (b), EoS curves are also significantly different as shown in Fig. ~\ref{fig:eos} (a). For the rest of REN's, EoS tests are presented in Supplementary materials (see section \ref{sec:appendix}). By taking the AE equilibrium lattice constant as reference, we can calculate \% errors of lattice constants as depicted in Fig. ~\ref{fig:eos} (b). Compared to PBE functional's typical overestimation\cite{overest} of lattice constants by about 1-2 \%, we believe that there is no need to reduce the errors in Fig. ~\ref{fig:eos} (b).

\section{Suggested Hubbard U values} \label{sec:match}

\begin{figure*}
\includegraphics[width=17cm]{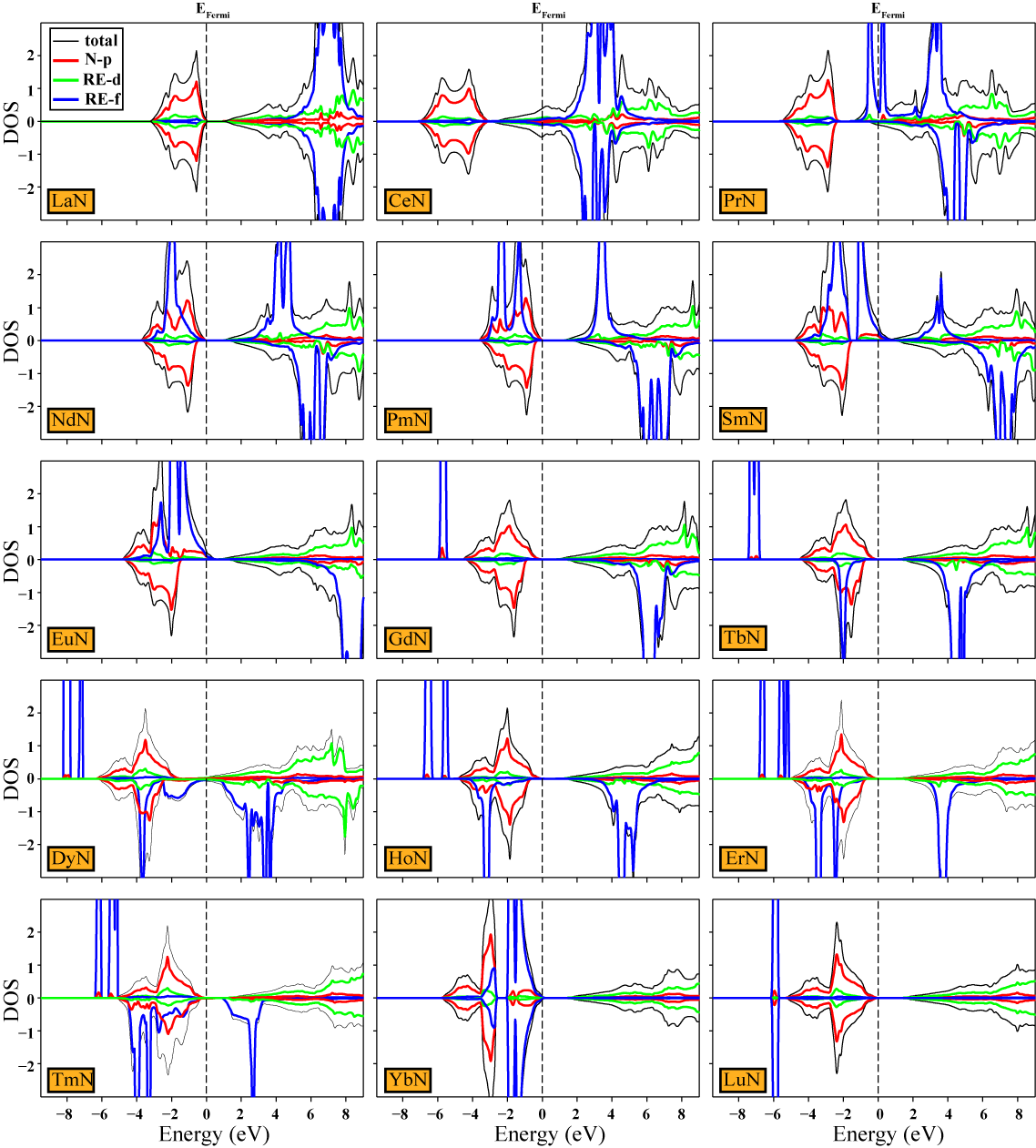}
\caption{(Color online) Total and projected density of states (DOS) of REN's calculated using all-electron hybrid functional (YS-PBE0) FLAPW+LO method. Positive (negative) y-axis corresponds to DOS of spin-up (spin-down) states. Zero of the energy was set to Fermi energy (E$_{Fermi}$).}
\label{fig:wHSE}
\end{figure*}

For the case of Cerium, J.L.F. Da Silva \textit{et al.}\cite{kresse_ce} and P.J. Hay \textit{et al.}\cite{hse_ce}  showed that hybrid functionals correctly predict $Ce_{2}O_{3}$ to be an insulator as opposed to the ferromagnetic metal predicted by the LDA and GGA. Calculated band gaps are also shown to be in close agreement with respect to available experimental data. For the case of GdN, M. Schlipf \textit{et al.}\cite{blu_hse} showed that FLAPW hybrid functional calculations attain good agreement with experimental data for band transitions, magnetic moments, and the Curie temperatures. 

In Fig. ~\ref{fig:wHSE}, we present hybrid functional (YS-PBE0) density of states (DOS) results for all REN's (La-Lu). As compared to previous DOS calculations in Fig. ~\ref{fig:000}, the YS-PBE0 functional yields significantly different DOS. RE 4\textit{f}-states, which usually tend to lie near the Fermi level in PBE calculations, get more localized and move away from the Fermi level. Especially in LaN, GdN, TbN, HoN, ErN, TmN, and LuN f-states are represented by distinct isolated peaks in the DOS. For CeN, PrN, SmN, EuN and YbN, f-states are still close to the Fermi level.

In section ~\ref{sec:comp}, we showed that DOS and EoS for all REN produced using our generated PAW datasets compare very well with those produced using FLAPW-LO. However, hybrid functional results differ significantly from standard PBE results. Assuming that hybrid functional results are more reliable, the aim of this section is to provide suggested Hubbard U values for prospective users of RE PAW datasets that we generated. Of course it is always desirable to use hybrid functionals for any DFT calculation involving RE elements. However, in most cases this can be challenging. From our experience, a typical YS-PBE0 calculation on a REN can be 10$^2$ to 10$^3$ times more time consuming than a standard PBE calculation. With primitive cells containing 10-20 atoms, it is challenging even to start a YS-PBE0 calculation due to memory requirements. Another issue is related with the implementation of hybrid functionals into open-source density functional simulation environments. For example, at the time this paper is written, it is not possible to perform a PAW hybrid functional calculation using current version of QUANTUM ESPRESSO (QE 5.03). Therefore, it is desirable to perform  DFT+U calculations that reproduce as closely as possible results from hybrid functional calculations. DFT+U calculations add little workload to standard DFT calculations. Sometimes U can be chosen to reproduce some well-established experimental spectroscopic data, which should in principle, be reproduced also by a hybrid functional calculation or using U values determined using linear response.\cite{matteo} In this section, we try to define U values for some RE elements to allow us perform inexpensive DFT+U simulations giving results similar to YS-PBE0 calculations. These U values are suitable for trivalent ions in the highest spin state. In principle, U varies with interatomic distances, atomic configurations, etc. However, these are small variations compared with those produced by different valence or spin states.\cite{han1,han2,han3}

\begin{figure}
\includegraphics[width=7.5cm]{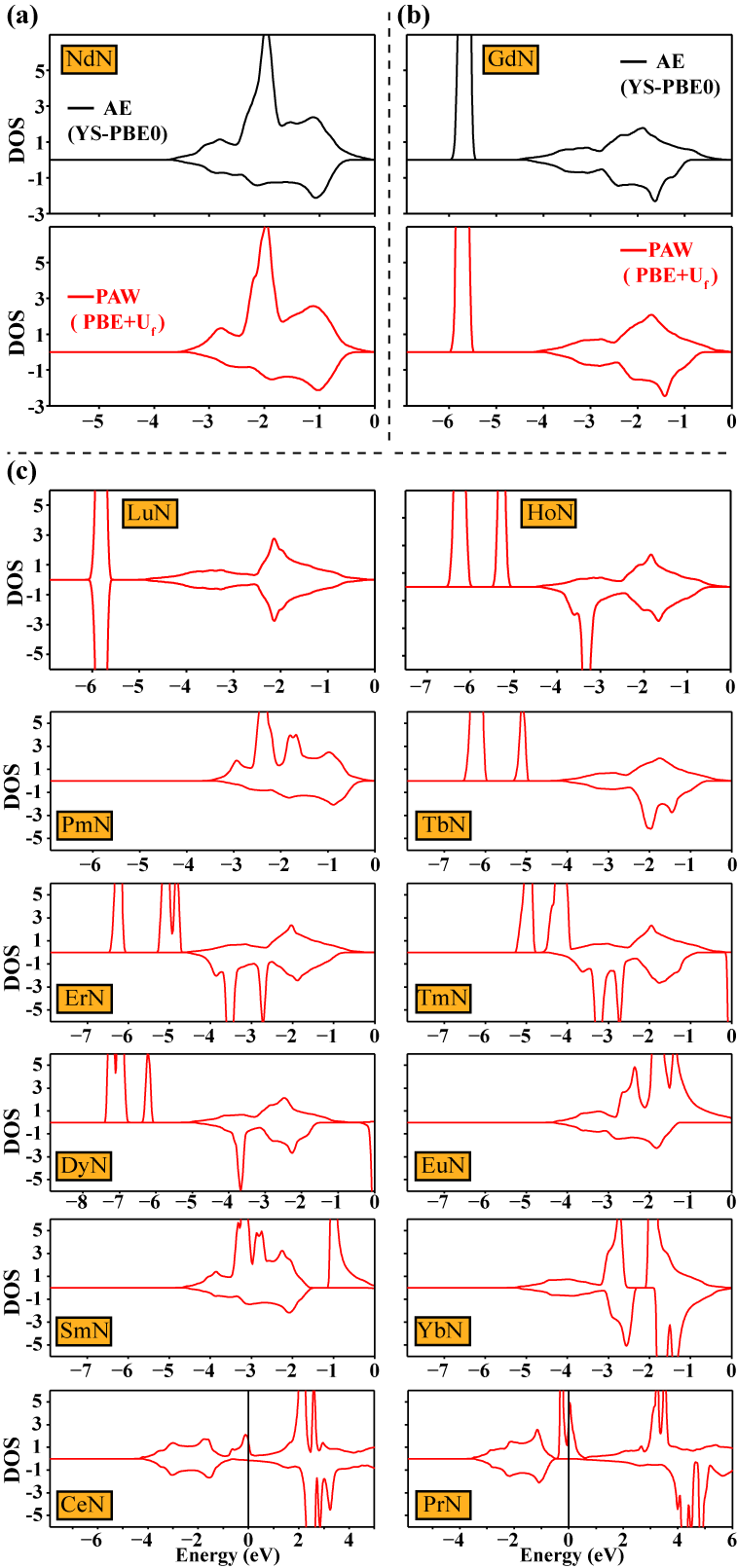}
\caption{(Color online) (a) Valance total density of states of NdN calculated with YS-PBE0 (upper) and PAW PBE+U$_f$  (lower) method. U$_f$=3.1 eV was used to match YS-PBE0 DOS. Zero of the energy was set to Fermi energy (E$_{Fermi}$). (b) Same as (a) for GdN. U$_f$=4.6 eV was used to match YS-PBE0 DOS. (c) PAW PBE+U$_f$ calculations for the rest of REN's. Their YS-PBE0 counterparts are presented in Fig. ~\ref{fig:wHSE} and the list of suggested U$_f$ values are presented in Table \ref{table:Us}. }
\label{fig:hse}
\end{figure}

In the upper panel of Fig. ~\ref{fig:hse} (a), valance DOS of NdN calculated by YS-PBE0 functional is shown. The peak due to f-states is located around -2 eV. This YS-PBE0 DOS is considerably different from the PBE DOS shown in Fig. ~\ref{fig:000} (e), but can be well reproduced with the DFT+U method using U$_f$  = 3.1 eV for Nd (Fig. ~\ref{fig:hse} (a)). Similarly good valence DOS agreement can be obtained with U$_f$ = 4.6 eV for Gd f-states in GdN, and U$_f$ = 4.6 for Lu f-states in LuN (see Fig. ~\ref{fig:hse} (b) and (c)). For HoN, there are three distinct f-peaks in the valance region and it is more difficult to obtain an exact match between YS-PBE0 and PBE+U DOS. However, one can obtain a fairly good agreement for U$_f$=4.95 eV. In Fig. ~\ref{fig:hse} (c), we present PBE+U$_f$ DOS results that resemble their YS-PBE0 counterparts shown in Fig. ~\ref{fig:wHSE} for the remaining REN's. The list of suggested U$_f$ values are presented in Table \ref{table:Us}. For YbN, CeN and PrN we are unable to find a U$_f$ value that closely match YS-PBE0 DOS. These cases are more subtle because of their orbital occupancies and non-zero J values (not considered in this study) are probably needed. Nevertheless, U$_f$  values provided here for trivalent high spin RE ions could be used in cases where no other alternative is available.

\begin{figure}
\includegraphics[width=8cm]{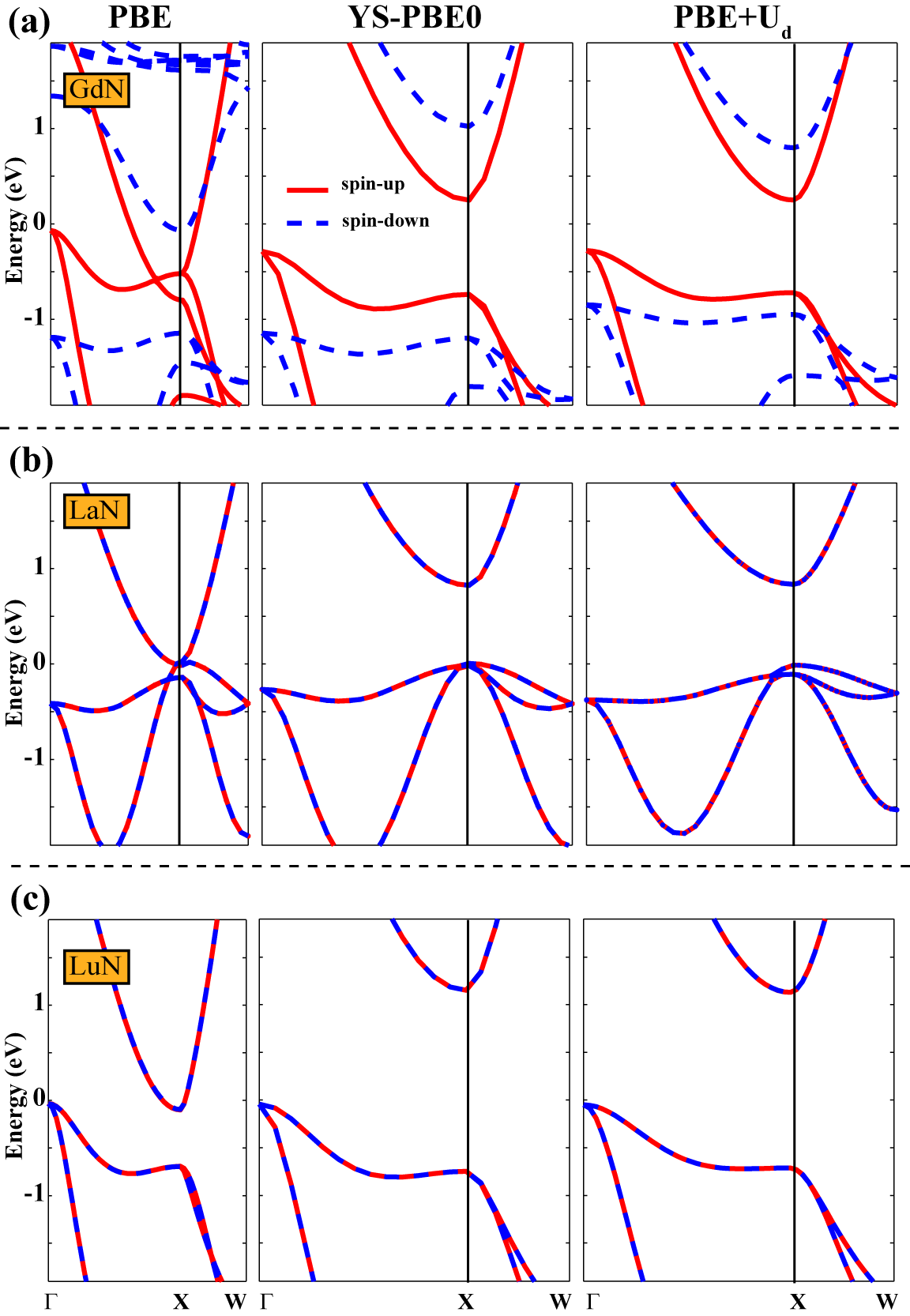}
\caption{(Color online) Band structure of GdN calculated with PBE, YS-PBE0 and PBE+U$_d$ methods along $\Gamma$-X-W. Continuous and dashed lines correspond to spin-up and spin-down bands respectively. U$_d$=9.5 eV was used to match YS-PBE0 band gap. (b) Same as (a) for LaN. U$_d$=9.0 eV was used to match the YS-PBE0 band gap. (c) Same as (a) for LaN. U$_d$=8.2 eV was used to match YS-PBE0 band gap.}
\label{fig:hse:band}
\end{figure}

Similar to f-orbitals, RE 5\textit{d} orbitals, which constitute the low energy conduction bands of REN's, are also problematic in standard DFT. The first panel in Fig. ~\ref{fig:hse:band} (a) shows the half-metal PBE band structure of GdN. However, experimental studies assert that GdN has a band gap around 1.3 eV\cite{gdn_exp1}, which is taken as the average of majority and minority band gaps. The middle panel in Fig. ~\ref{fig:hse:band} shows the YS-PBE0 bands with an average band gap of 1.40 eV, which is quite close to the experimental value. By taking YS-PBE0 band structure as a reference, one can obtain a similar band structure by applying U$_d$ = 9.5 eV to Gd-d orbitals as shown on the third panel of Fig. ~\ref{fig:hse:band} (a). This procedure can be repeated for LaN and LuN as shown in Fig. ~\ref{fig:hse:band} (b) and (c). U$_d$ = 9.0 (8.2) eV can be used to reproduce the YS-PBE0 band gap of LaN (LuN). The list of suggested U$_d$ values are shown in Table \ref{table:Us}. Due to the complex nature of states near Fermi level, we cannot safely determine U$_d$ values for some of REN's. In average, 8.5 eV U$_d$  is expect to work for most of them.

\begin{table}
\caption{Suggested Hubbard U$_f$ and U$_d$ values  (see text).}
\label{table:Us} \centering{}
\begin{tabular}{|c|c|c|c|}
\hline
Element & U$_f$ (eV)& U$_d$ (eV)
\tabularnewline
\hline
\hline
La  & 5.5 & 9.0    \tabularnewline
\hline
Ce  & 2.5 & -  \tabularnewline
\hline
Pr  & 4.0 & -   \tabularnewline
\hline
Nd  & 3.1 & 8.5   \tabularnewline
\hline
Pm  & 3.4 & 8.1   \tabularnewline
\hline
Sm  & 3.3 & -   \tabularnewline
\hline
Eu  & 3.0 & -   \tabularnewline
\hline
Gd  & 4.6 & 9.5   \tabularnewline
\hline
Tb  & 5.0 & 9.0   \tabularnewline
\hline
Dy  & 5.0 & -   \tabularnewline
\hline
Ho  & 4.9 & 8.4  \tabularnewline
\hline
Er  & 4.2 & -   \tabularnewline
\hline
Tm  & 4.8 & -   \tabularnewline
\hline
Yb  & 3.0 & 8.0   \tabularnewline
\hline
Lu  & 5.5 & 8.2   \tabularnewline
\hline
\hline
\end{tabular}
\end{table}

\section{Conclusions} \label{sec:conc}

In conclusion, we have generated accurate and publicly available PAW datasets for RE elements (La-Lu) which can be used readily with open-source QUANTUM ESPRESSO and ABINIT DFT simulation packages. Similar to common trivalent RE ions found in nature, solid state tests and optimizations of PAW datasets were performed on rare-earth nitrides. We find that our optimized PAW datasets yield almost identical results to highly accurate all-electron full-potential linearized augmented plane-wave plus local orbital (FLAPW+LO) calculations. All-electron hybrid functional calculations (YS-PBE0) were carried out to overcome limitations of standard PBE calculations and to be used as reference for the determination of Hubbard U values for PBE+U calculations. PBE results tend to place f-states very close to the Fermi level and 2.5 to 5.5 eV Hubbard U values are required to place f-states in positions similar to those produced by the YS-PBE0 hybrid functional. Nonzero Hubbard U value ($\sim8.5$ eV) on empty RE d-orbitals was also shown necessary to open a band gap of some REN's, which is consistent with experimental findings. We believe that these new PAW datasets - if used with suggested Hubbard U values (for trivalent high-spin ions) - will allow further studies on rare-earth materials.

\section{Acknowledgments}
This work was supported primarily by NSF/DMR through the University of Minnesota MRSEC. RMW was also supported by NSF/EAR 1319361. We acknowledge fruitful discussions with Dr. Koichiro Umemoto, Prof. Natalie Holzwarth and Prof. David Vanderbilt. Computations were performed at the Minnesota Supercomputer Institute (MSI).

\section{Appendix A. Access to generated PAW datasets} \label{sec:access}
ATOMPAW input files used in the generation of LDA and GGA PAW datasets and corresponding output potential files which can be used with ABINIT and QUANTUM ESPRESSO are publicly available at http://www.vlab.msi.umn.edu/resources/repaw/index.shtml.

\section{Appendix B. Supplementary material} \label{sec:appendix}
Supplementary material associated with this article can be found in the
online version at http://dx.doi.org/10.1016/j.commatsci.2014.07.030.


\begin{thebibliography}{15}
\bibitem{Pm}                 Promethium (Pm) is a radioactive element and experimental lattice constant is not available for PmN. We take the average of NdN and SmN lattice constants for PmN.
\end{thebibliography}
\end{document}